\documentclass{article} 
\usepackage{graphicx}        
\begin{document}
\centerline{\bf Love kills Penna Ageing Model}
\medskip

Dietrich Stauffer* and Stanis{\l}aw Cebrat
\medskip

Department of Genomics, 
Wroc{\l}aw University, ul. Przybyszewskiego 63/77, 51-148 Wroc{\l}aw, Poland
\medskip

* visiting from Institute for Theoretical Physics, Cologne University,
D-50923 K\"oln, Euroland
\bigskip

Love may have been a female invention long ago, when the brain of genus homo
became large and required more food. Thus the father was needed to help feed
mother and baby. Pekalski \cite{pek} assumed mate selection to be governed by a Major
Histocompatibility Complex MHC. In the same spirit now the sexual Penna 
ageing model is modified to include mate selection by an additional string of
32 bits for each individual, unrelated to the two usual bit-strings of length 
$L$ containing the age-relevant genome. Initially, the additional love 
bit-string is chosen randomly, different for each individual. This model is 
therefore more complicated than the gamete recognition of Cebrat and 
Stauffer \cite{ceb} based on one bit.

During the at most 20 attempts per iteration of a female to find a suitable 
male partner, the two new ``love'' bit-strings of the male and the female
which should be similar, are compared bit by bit. If the number of differences
is larger than a universal love limit between 0 and 32, the male is rejected.
Thus with too stringent requirements for love, the female will often not
find a suitable partner within the allowed 20 attempts, stay single during 
this iteration, thus reduce the total number of new babies, and finally lead 
the population to extinction, as indicated for Germany by present trends. 
(Sons get the love bit-string from their father, and daughters from their 
mother.) 

We start with the standard Penna model \cite{penna} with 
$L=64$, a minimum reproduction age $R = 16$, $T = 3$ active mutations kill, 
$B = 2$ births are attempted per female and iteration, $d = 8$ loci are 
dominant, both males and females suffer from one (deleterious) mutation per 
bitstring and iteration, Verhulst deaths are applied to births only with a
carrying capacity $K$ up to 30 million. If the male partner can differ in at 
most $6 \pm 1$ bit pairs (i.e. they agree in at least ($32-6=26$ bit pairs) to
remain 
suitable, the population survives; with the limit one bit less, the population
dies down. For $L = 1024, \, R = 256$ and $K$ up to 3 million instead, 
survival requires $5 \pm 1$ bits. If observation time would go to infinity at 
a fixed population size and a small love limit,
we might get an Eve effect for the love bit-strings: All surviving males
have one string and all surviving females the same or the complementary
bit-string. However, if our populations vanish they do so rather quickly as
shown in our figure, and prevent this Eve effect. (If love starts only after
10,000 iterations, close to equilibrium, the limits are higher.)

Love entails the danger of monogamy. Assuming that a loving couple stays 
together until death, but then can remarry, we find the threshold to decrease
from about six bits for $K$ = thousand to about zero or one bit for $K$ = 30 
million.

\begin{figure}[hbt]
\begin{center}
\includegraphics[angle=-90,scale=0.32]{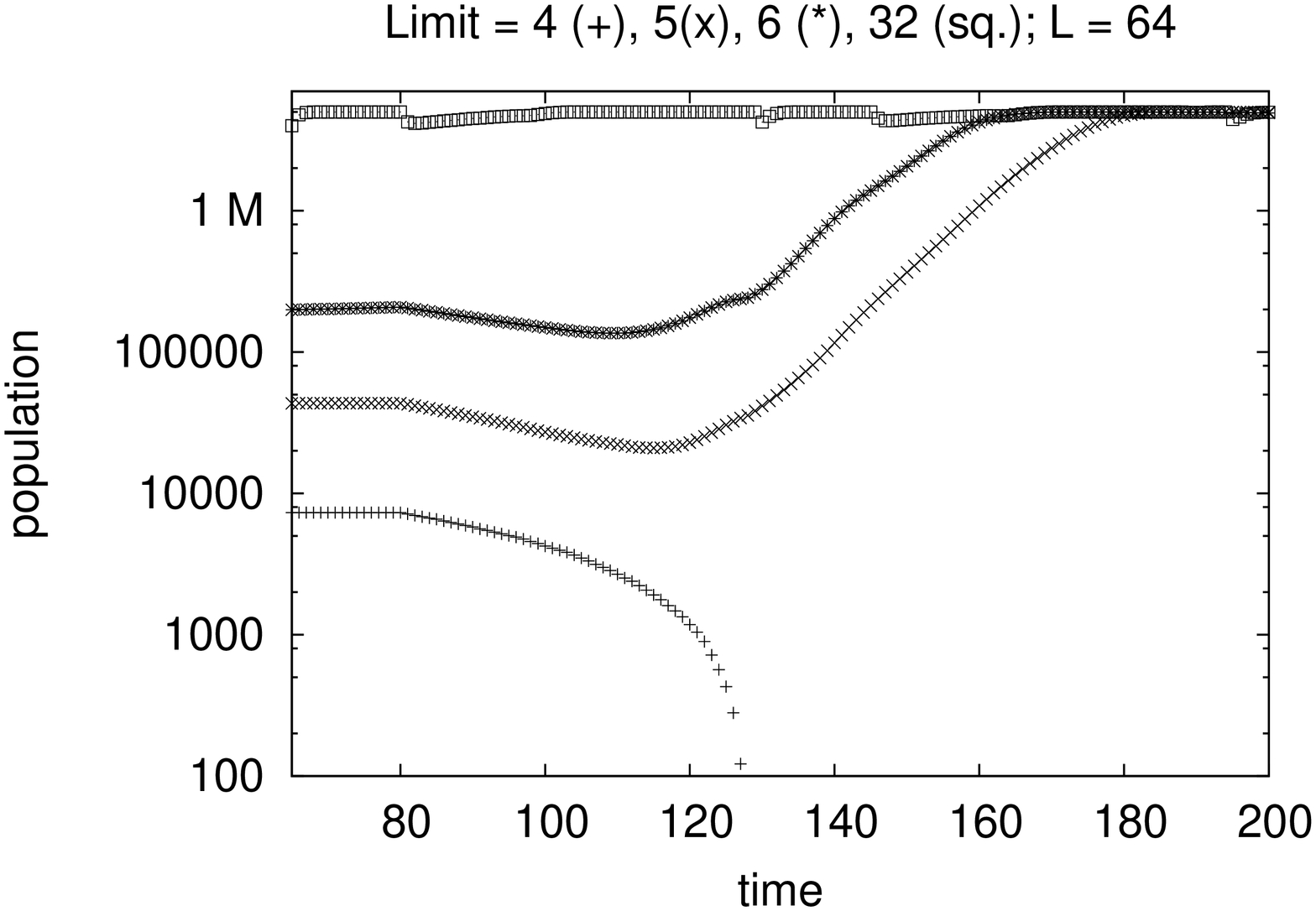}
\includegraphics[angle=-90,scale=0.32]{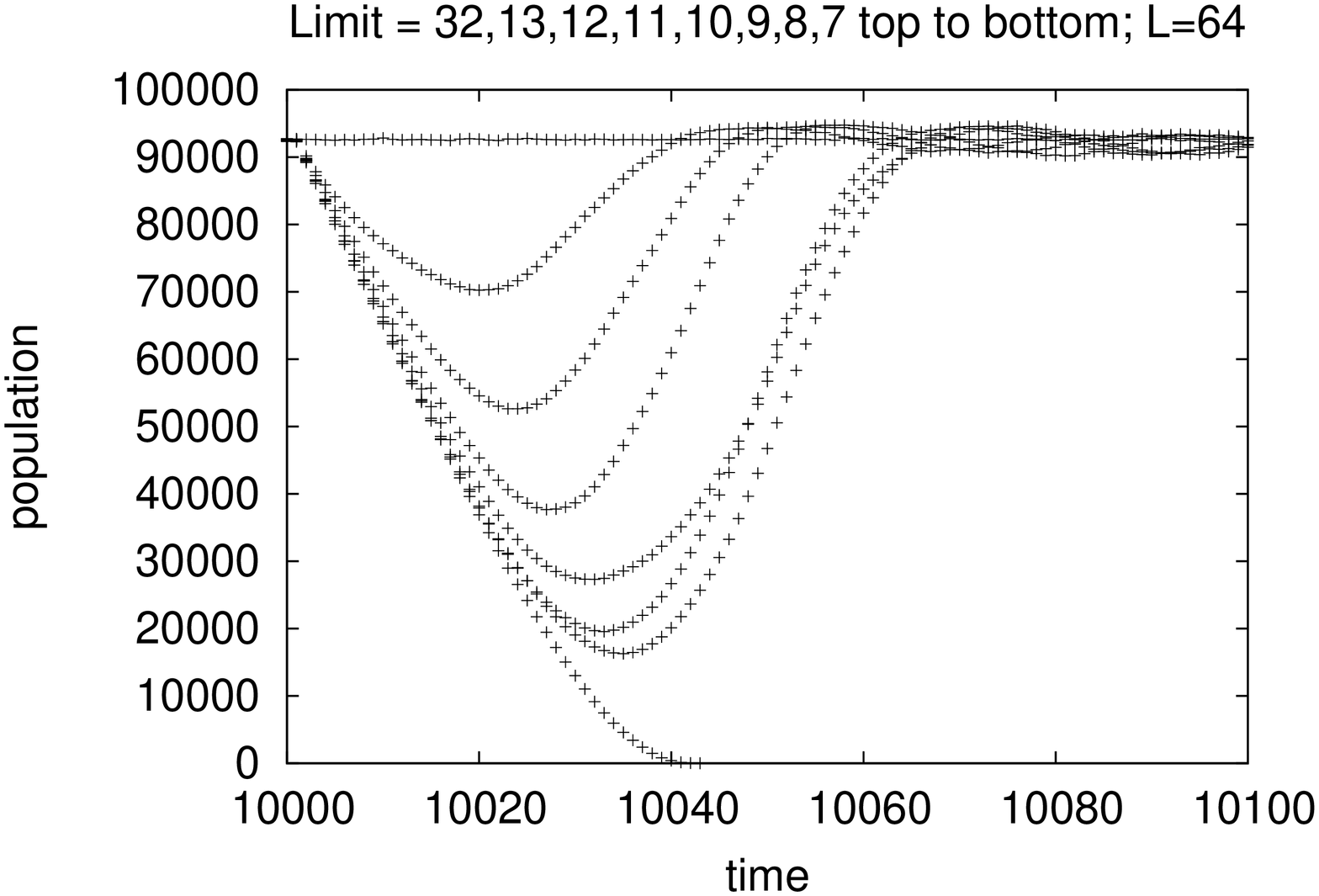}
\end{center}
\caption{a: Sexual Penna model with love (lower curves) and without love 
restriction (upper squares): Extinction threatens with too narrow requirements
for love. b: Love is switched on after 10,000 iterations.
}
\end{figure}

\end{document}